\documentclass[draft]{agujournal2019}
\usepackage{url} 
\usepackage{lineno}
\usepackage[finalnew]{trackchanges} 
\usepackage{soul}
\draftfalse

\journalname{JGR: Space Physics}

\begin{document}


\title{Understanding and Modeling the Dynamics of Storm-time Atmospheric Neutral Density using Random Forests}


\authors{K. Murphy\affil{1,2}, A.J. Halford\affil{3}, V. Liu\affil{4}, J. Klenzing\affil{3}, J. Smith\affil{5}, K. Garcia-Sage\affil{3}, J. Pettit\affil{3}, I. J. Rae\affil{2} }

\affiliation{1}{Lakehead University, Thunder Bay, Ontario, Canada}
\affiliation{2}{Northumbria University, Newcastle upon Tyne, UK}
\affiliation{3}{NASA Goddard Space Flight Center, Maryland, USA.}
\affiliation{4}{Johns Hopkins University, Maryland, USA}
\affiliation{5}{Catholic University of America, DC, USA}

\correspondingauthor{Kyle Murphy}{kylemurphy.spacephys@gmail.com}

\begin{keypoints}
\item The differences in storm-time vs quiet-time density variations are the result of geomagnetic processes
\item Random Forest models capture storm-time atmospheric density variations with correlation squared exceeding 0.8.
\item Model performance is maximized when using high-cadence solar and geomagnetic data capable of capturing rapid storm-time changes
\end{keypoints}

%
%

%
%


\begin{abstract}

Atmospheric neutral density is a crucial component to accurately predict and track the motion of satellites. During periods of elevated solar and geomagnetic activity atmospheric neutral density becomes highly variable and dynamic. This variability and enhanced dynamics make it difficult to accurately model neutral density leading to increased errors which propagate from neutral density models through to orbit propagation models. In this paper we investigate the dynamics of neutral density during geomagnetic storms. We use a combination of solar and geomagnetic variables to develop three Random Forest machine learning models of neutral density. These models are based on (1) slow solar indices, (2) high cadence solar irradiance, and (3) combined high-cadence solar irradiance and geomagnetic indices. Each model is validated using an out-of-sample dataset using analysis of residuals and typical metrics. During quiet-times, all three models perform well; however, during geomagnetic storms, the combined high cadence solar iradiance/geomagnetic model performs significantly better than the models based solely on solar activity. \add[KRM]{The combined model capturing an additional 10\% in the variability of density and having an error up to six times smaller during geomagnetic storms then the solar models}. Overall, this work demonstrates the importance of including geomagnetic activity in the modeling of atmospheric density and serves as a proof of concept for using machine learning algorithms to model, and in the future forecast atmospheric density for operational use.

\end{abstract}

\section{Plain Language Summary}

Even though satellites are in space they still experience drag or friction as they fly through what little atmosphere there is along their orbit. This drag causes satellite orbits to decay overtime. During periods of enhanced space weather, such as solar flares or geomagnetic storms, this drag increases which can make it difficult to track and predict the motion of satellites. In this work we develop a new model for atmospheric density, they key contributor to satellite drag, utilizing combined solar and geomagnetic data. This new model  performs better than models based only on solar data especially during geomagnetic storms, periods of extreme space weather. Overall this work highlights the importance of near Earth processes in enhancing satellite drag during geomagnetic storms.

\section{Introduction}

The Sun-Earth system is a highly dynamic and extremely coupled environment. The physical processes coupling and driving the dynamics of the solar wind-magnetosphere-ionosphere-thermosphere system can affect critical ground- and space-based technological infrastructure \cite<e.g.,>[]{Blake2016, Cassak2017, Morley2020, Licata2020, Bodeau2021, Chakraborty2022, Klenzing2023}. For example, intense ionospheric currents can cripple ground-based power systems \cite{Oughton2018, Cid2020}, highly energetic electrons in the Earth’s radiation belt can lead to satellite anomalies and even complete failures \cite<e.g.,>[]{Green2017, Berthoud2022}, ionospheric irregularities can affect GNSS signal propagation and hinder communications and ionosphere-thermosphere heating can increase atmospheric density leading to increased satellite drag, shorter satellite lifetimes, and even the complete loss of satellite infrastructure \cite<e.g.,>[]{Carter2020, Thayer2021, Fang2022, Carter2023}. Mitigating these effects requires high-fidelity models capable of capturing the dynamic processes adversely affecting technological infrastructure \cite<e.g., >[]{Zhang2018, Sutton2018, Licata2022, Ponder2023}. Such models can be used in operational environments to aid stakeholders in making key decisions to protect various technologies and infrastructure. 

Of the processes and impacts described above, accurately modeling atmospheric density is particularly interesting to space weather forecasters and stakeholders \cite<e.g.,>[]{Berger2020}. Robust models capable of forecasting the dynamics of atmospheric density increase the fidelity of orbit propagation, allowing for more accurate tracking of satellites and debris in low-Earth orbits (LEO). As the number of satellites and orbital debris in LEO continues to grow, such tracking will become increasingly important as it is key for identifying potential collisions and defining collision avoidance maneuvers.   

The density of the ionosphere-thermosphere system is driven by a combination of external forcing from the Sun \cite{Lilensten2008} and the magnetosphere \cite{Knipp2004} and internal processes in the lower atmosphere \cite{Liu2016}. Generally, Extreme Ultra Violat (EUV) and Ultra Violet (UV) emissions from the sun are the dominant sources of the dynamics of atmospheric density driving variations on time scales of the order of several days, a solar rotation, and over a solar cycle, \cite<e.g., >[]{Qian2011}. However, during geomagnetic storms, energy input and forcing from the magnetosphere are the dominant drivers of the dynamics of atmospheric density  \cite<e.g., >[]{Knipp2004, Zesta2019} driving rapid enhancements on time scales from  several hours (and potentially shorter), which can last for several days \cite<e.g., >[]{Liu2005, Oliveira2019}. During periods of quiet solar and geomagnetic activity, internal processes such as atmospheric waves can contribute significantly to the dynamics and redistribution of atmospheric density \cite{Liu2016}. Together, solar, magnetospheric, and internal forcing drive a complex set of dynamics in the ionosphere-thermosphere system across a broad range of temporal scales, spatial scales, and spatial regions including global and local variations with latitude, longitude, and altitude. In this work, we focus on quantifying and modeling the storm-time dynamics of atmospheric density, a key challenge for space weather modeling and forecasting.

During geomagnetic storms, energy input into the ionosphere-thermosphere system in the form of field-aligned currents \cite<e.g.,>{Luhr2004}, particle precipitation \cite[e.g.,]{Deng2013}, and joule heating \cite<e.g.,>{Kim2006, Wang2022} can lead to rapid changes in atmospheric density ranging from 50-800\% \cite{Forbes1996, Liu2016, Oliveira2019}. During storms, density enhancements are first observed at higher magnetic latitudes, quickly migrating to lower latitudes over a few hours \cite{Zesta2019}. They are typically scaled with the storm's size, such that the largest changes in density are observed during the most extreme geomagnetic storms \cite{Oliveira2019}. In general, existing thermospheric models have difficulty capturing the dynamic spatiotemporal evolution of atmospheric density during geomagnetic storms. This is partly because models' cadence and spatial resolution are insufficient to capture storm-time thermospheric dynamics \cite<e.g.,>{Bruinsma2018} and typically underestimate magnetospheric forcing \cite<e.g.,>{Lu2023} . Recent model developments incorporating higher cadence solar and geomagnetic inputs have helped address this \cite<e.g.,>{Bruinsma2021}; however, additional research and model development are required to accurately simulate storm-time atmospheric density and, more importantly, accurately forecast atmospheric density for use in operational space weather.

In this work, we investigate the solar and geomagnetic drivers of atmospheric density changes during geomagnetic storms (storm-time) and geomagnetic quiet periods (quiet-time). Building on this, we develop three Random Forest machine learning models of atmospheric density, one using low-cadence solar indices, a second using high-cadence solar spectra, and a third using combined high-cadence solar spectra and geomagnetic indices. During quiet-times all three models perform well; however, during geomagnetic storms the combined high-cadence solar/geomagnetic model performs significantly better than the models based solely on solar activity. Overall, this work demonstrates the importance of accurately capturing geomagnetic activity in the modeling of atmospheric density and serves as a proof of concept for using machine learning algorithms for hindcasting, nowcasting, and eventually forecasting atmospheric density for operational use.

In the subsequent sections, we detail the datasets used to study and quantify the drivers of storm- and quiet-time atmospheric density. We then detail the analysis of the relation of solar indices, solar irradiance, and geomagnetic activity to atmospheric density during storms and quiet times. Following this, we describe the development of three Random Forest models, the quantification of the performance of each model, as well as the relative importance of each feature used within each model. The three models are then compared during four increasingly active periods of geomagnetic activity and the model errors are quantified as a function of storm-time and geomagnetic activity. These results are summarized, and we conclude with future work and directions for extending the machine learning models developed here. 

\section{Datasets}

In this work, we use a combination of in-situ neutral density measurements from the dual Gravity Recovery and Climate Experiment (GRACE A and B) and the CHAllenging Minisatellite Payload (CHAMP) satellites, solar irradiance data from the Flare Irradiance Spectral Model 2 \cite<FISM2,>{Chamberlin2020, Solomon2005}, solar indices used in atmospheric models \cite{Bowman2008, Tobiska2008}, and solar wind data and geomagnetic indices from the OMNI dataset \cite{King2005}.  

CHAMP was launched in 2000 \cite{Reigber2002}, and shortly after, the dual spacecraft GRACE mission was launched in 2002 \cite{Wahr2004}. Both missions were launched into near-polar low-Earth circular orbits to altitudes of a radius of \~460 km and  \~500 km, respectively. This work utilizes neutral densities derived from high-precision accelerometers onboard each spacecraft \cite{Sutton2005, Sutton2009}. The along-track density is then normalized to an altitude of 400 km using the empirical global reference atmosphere model NRLMSIS-00 \cite{Picone2002}. This normalization allows us to remove altitude as an independent variable when characterizing the dynamics of atmospheric neutral density as a function of solar and geomagnetic activity. The neutral density data is available through the University of Colorado Boulder Space Weather Data Portal. 

The FISM2 and solar indices are used as proxies for solar energy input to the upper atmosphere, leading to variations in neutral density. The FISM2 flare dataset is an empirical model of solar spectral irradiance from 0.01-190 nm in 0.1 nm spectral bins with a 60-second cadence \cite{Chamberlin2020}. Here, we use a reduced FISM2 flare dataset comprised of 23 spectral bands \cite{Solomon2005}. These bands are commonly used in global time-dependent thermosphere-ionosphere models as they reduce both the dimensionality of model input data and the computational overhead of models without excessive loss of model accuracy \cite{Solomon2005}. The solar indices used in this study are the F10, S10, M10, and Y10 indices and their 81-day centered averages.  These indices are used in the semi-empirical thermosphere model JB2008 and have a 24-hour cadence \cite{Bowman2008}. The indices correspond to different portions of the solar irradiance spectrum spanning UV to EUV to X-Ray; a detailed description of the indices can be found in Tobiska et al. \citeyear{Tobiska2008}. OMNI data from NASA's Space Physics Data Facility provide solar wind and geomagnetic observations \cite{King2005}. These observations can be related to physical processes that drive magnetosphere and ionosphere dynamics, including geomagnetic storms, substorms, and enhanced precipitation. Here, we use OMNI to quantify the solar wind and geomagnetic drivers of elevated neutral atmospheric density during both quiet and storm-time conditions. Finally, we use a database of geomagnetic storms between 2002-2012, inclusive, so that the datasets can be separated by quiet-time, storm-time, and storm phase (main or recovery). The database of storms is developed using the methodology outlined in \citeA{Murphy2018, Murphy2020}. 

The CHAMP, GRACE, FISM2, solar, and OMNI data have varying cadences and time stamps. The CHAMP and GRACE data have a cadence of $\sim$50s, the solar indices and FISM2 flare data have a cadence of 24h and 60s (respectively), and the OMNI data has a 5m cadence. These datasets must have similar cadences and timestamps to perform any relational study between atmospheric density and solar and geomagnetic activity. This is typically achieved by interpolating the datasets to a common abscissa. However, interpolation routines perform poorly for data that rapidly vary, in this case, the CHAMP and GRACE density datasets. Here, we use a nearest-neighbor approach matching the density and solar datasets to the time stamps of the OMNI data. Each satellite is independently matched to the driver datasets (solar, FISM2, OMNI), creating a total of three databases: GRACE-A, GRACE-B, and CHAMP. \add[KRM]{When matching timestamps all bad data are dropped (e.g., missing data as identified by the datasets metadata or NaN's) and} the time stamps of each database are  tagged as either quiet or storm-times, with storm-time further tagged by storm phase using the storm list \citeA{Murphy2018, Murphy2020}.  These databases are used to investigate the dynamics of neutral density as a function of solar and geomagnetic activity and subsequently develop a random forest model of neutral density. In developing the random forest models, the GRACE-B database is used to train the model, and the GRACE-A and CHAMP databases are used as out-of-sample validations. \add[KRM]{Note, each of the three databases contain ~1 million discrete 5m time points and span nearly an entire solar cycle, 2002-2012. Over this period each satellite samples the entire range of latitudes and local times, spending over 10,000 minutes in area of $10^{\circ}$ latidue x 1hr MLT. Further, having the database span nearly an entire solar cycle ensures that both the solar and geomagnetic data cover a wide range of activity from quiet-times to storm-times. Together, the large size and coverage of the databases is sufficient for both the statistical and random forest analyses presented in subsequent sections.} 

\section{Atmospheric Density vs Solar and Geomagnetic Activity}

The dynamics of atmospheric neutral density resulting from solar and geomagnetic forcing is complex, responding to a combination of solar drivers and varying geomagnetic processes, including geomagnetic storms, substorms, field-aligned currents, and energetic particle precipitation. In this section, we investigate the dynamics of atmospheric density observed by the GRACE-B satellite as a function of quiet and storm-times, solar irradiance, solar wind activity, and geomagnetic activity.  This analysis is used to inform the development of a random forest model of neutral atmospheric density and identify key features (independent variables) for the model.

\begin{figure}
    \includegraphics{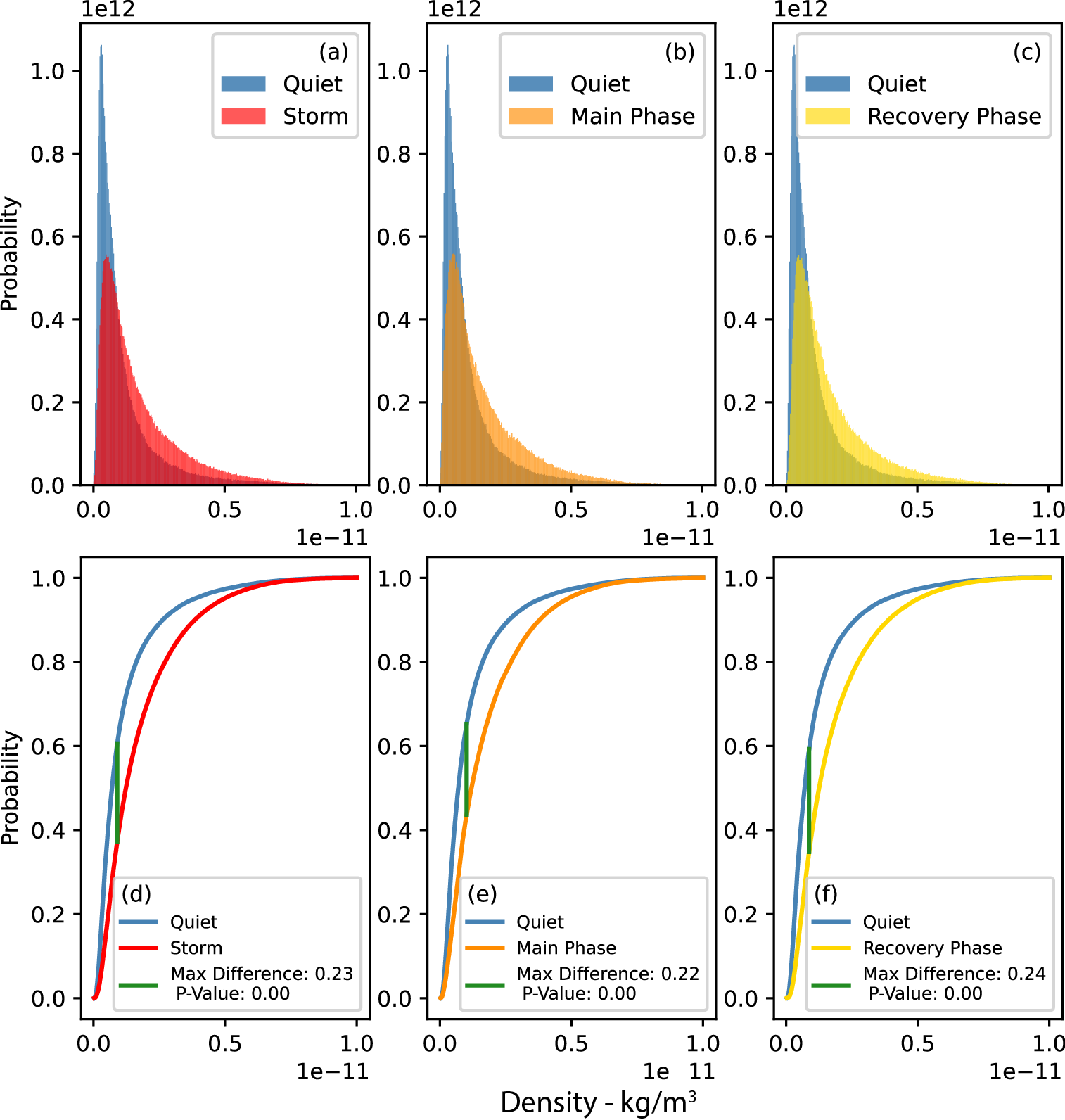}
    \caption{Top, probability distributions of atmospheric density at GRACE-B as a function of (a) quiet- and storm-times, (b) quiet-times and storm main phase, and (c) quiet times and storm recovery phase. Bottom, cumulative distributions of atmospheric density as a function of (d) quiet- and storm-times, (e) quiet-times and storm main phase, and (f) quiet-times and storm recovery phase. The green line in the cumulative distributions shows the maximum difference between each plot's quiet-time distribution and storm distribution. The KS statistic (Max Difference) and p-value are shown in the legend.} 
    \label{fig:1}
\end{figure}

Figure \ref{fig:1} shows the distribution of atmospheric neutral density at GRACE-B as a function of quiet-times and storm-times. The top row shows the probability distributions (integrates to unity), and the bottom row shows the cumulative distributions; quiet times are shown in blue, storm-times in red, main phase in orange, and recovery phase in yellow. Evident in both the probability and cumulative distributions is that atmospheric density is enhanced during geomagnetic storms. The probability distributions (a-c) show that quiet-time densities are concentrated at lower values with a sharp peak and a rapid decay. In contrast, storm-time densities have smaller peaks at low densities and a clear enhancement in atmospheric density as compared to quiet times. The cumulative distributions (d-e) show the same trend; quiet-time densities rapidly reach the asymptotic limit of 1 at significantly lower densities than storm-time and storm phase such that higher densities are observed more often during storms than quiet times.

The green lines on the cumulative distributions (d-e) show the maximum difference between the quiet and storm-time distributions. This value can be used in the Kolmogorov-Smirnov (KS) test to determine whether two datasets are consistent with being drawn from the same distribution. Here, the KS test is used to quantify that the quiet-time density distribution is statistically different from the storm-time distribution. In the case of storms, the difference between the quiet-time and storm-time and storm phase densities is large, and the p-value is negligibly small (in part due to the large size of the datasets). This suggests that the distribution of storm-time atmospheric density is statistically and significantly different than the quiet-time distribution of atmospheric density. The values for both the KS-statistic and p-value are shown in the legend of Figure \ref{fig:1}.    

\begin{figure}
    \includegraphics{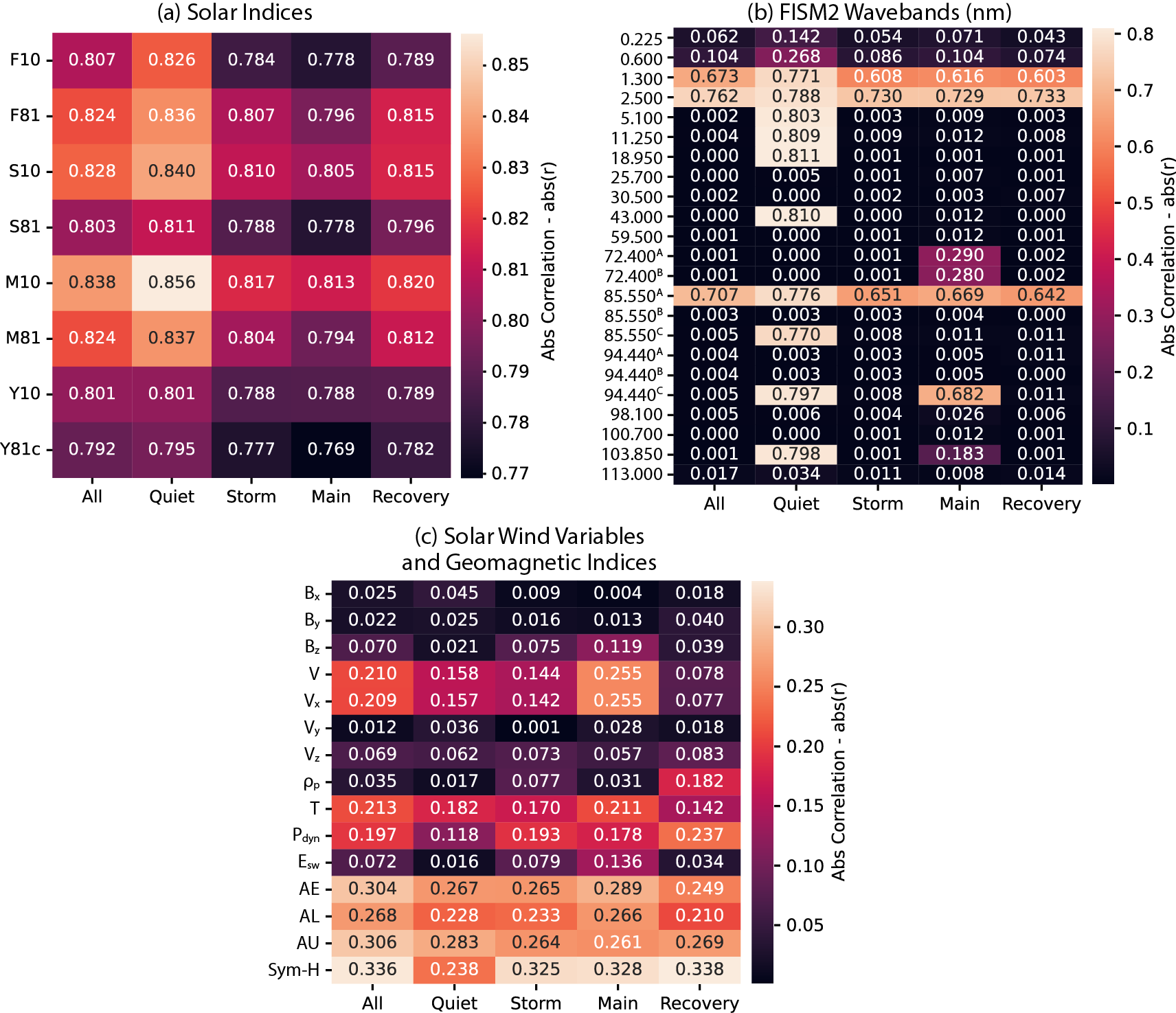}
    \caption{Correlation matrices for atmospheric density and (a) solar indices, (b) the FISM2 dataset and (c) solar wind data and geomagnetic indices (y-axis) as a function of all-time, quiet-time, storm-time and storm main and recovery phases (x-axis). The color and number indicate the absolute value of the correlation.}
    \label{fig:2}
\end{figure}

Expanding on the analysis shown in Figure \ref{fig:1} we use correlation matrices to investigate the solar and geomagnetic drivers of atmospheric density during quiet- and storm-times. Figure \ref{fig:2} (a-c) shows the correlation matrices of atmospheric density and solar indices, atmospheric density and the FISM2 dataset, and atmospheric density and the OMNI dataset (y-axis) as a function of all-times, quiet-times, storm-times, storm main phase, and storm recovery phase (x-axis). Figure \ref{fig:2} (a) shows that each solar index and its 81-day centered average correlate well with atmospheric density. The highest correlations are observed during quiet-times (0.80-0.85), while the lowest is during storm-times and specifically the main phase of storms (\~0.8). The correlations between atmospheric density and the M10 index are also generally the largest. Of note, the similarity in the correlations between the day of and 81-day centered average correlations suggest that solar driving is important for long term variations in atmospheric density on the order of several months, as opposed to short term variations on the order of hours and days, time-scales associated with geomagnetic storms.

The correlation between the 23 spectral bands from FISM2 and atmospheric density is shown in Figure \ref{fig:2} (b). Here there are clear differences between quiet and storm-times for several bands; several shorter wavelengths between \(1.300-18.950\; nm\) and select longer wavelengths \(85.550^C,\; 94.440^C,\; 103.850^C\; nm\) correlate well during quiet times, while the \(85.550^A\; nm\) correlates well during quiet and storm intervals. The correlations in these bands are comparable to those of the solar indices, the largest ranging from 0.77-0.81 during quiet-times and 0.65-0.73 during storm-times. Compared to panel (a), the results in panel (b) suggest that key spectral bands, 1.3, 2.5, and 85.5 nm, are important to the dynamics of atmospheric density during both quiet- and storm-times.

Finally, Figure \ref{fig:2} (c) shows the correlations of atmospheric density with solar wind variables and geomagnetic indices. These correlations are lower than those of both the solar indices and FISM2 wavebands. However, the correlations peak with auroral activity as measured by AE, AU, and AL, and storm activity as measured by Sym-H. The correlations in Figure \ref{fig:2} (c) are lowest for solar wind variables, the largest being 0.255 between solar wind velocity and main phase atmospheric density.

Taken together, Figures \ref{fig:1} and \ref{fig:2} demonstrate two key points. First, solar activity is the key factor controlling the background level of atmospheric neutral density, as demonstrated by the high correlations between solar indices and select spectral bands from the FISM2 database. Second, storm-time atmospheric neutral density has a different distribution than quiet times with increased higher density values and proportionally fewer lower density values. Previous research has suggested that the difference between storm-time and quiet-time densities is likely the result of increased geomagnetic activity leading to enhanced joule heating, field-aligned currents, and auroral precipitation during geomagnetic storms \cite{Knipp2004}. These results are used to develop and compare three random forest atmospheric density models. The goal is to develop a high-cadence model of atmospheric density capable of capturing both quiet- and storm-time dynamics.  

\section{Atmospheric Density and Random Forests}

Machine learning has proven to be a vital resource in Heliophysics. It has been used to develop models of magnetospheric dynamics \cite{Bortnik2018, Li2023}, in space weather forecasting \cite{Hua2022, Iong2022}, and as a tool for scientific discovery \cite{Camporeale2022}. Here, we utilize Random Forests, an ensemble machine-learning algorithm based on Decision Trees. In short, Decision Trees create a model that predicts the value of a target \add[KRM]{(or dependent variable)}, in this case, atmospheric density, by learning simple decision rules inferred from the data features \add[KRM]{(or independent variables)}, here solar indices and FISM2 and OMNI data. \add[KRM]{As a simple analogy, a decision tree is similar to an empirical model using binning or histogramming of a target (dependent variable) based on some independent variable, feature, or driver} \cite<e.g.,>{Gu2012}. \add[KRM]{However, a decision tree is more complex in that the bins, or decisions, are based on multiple independent variables or features. This additional complexity can increase the performance and accuracy of decision trees compared to simpler empirical models.} A Random Forest trains several Decision Trees, \add[KRM]{using an random sample of the input data to train each individual tree}. A final target value (the dependent variable) is then constructed by averaging the target from each Decision Tree. \add[KRM]{An added benefit of Random Forests and Decision Trees is that the feature and target data do not need to be normalized.}  

Random Forests are particularly useful in model development and model comparisons as there are several methods to investigate and quantify the relative importance of features within a model. Here, we use the mean decrease in accuracy (MDA, also referred to as permutation importance). The MDA measures the average decrease in accuracy when a feature vector is randomly shuffled. Randomly shuffling a feature decreases model accuracy, the larger the decrease in accuracy of the model the more important the feature is. In this methodology, a Random Forest model's accuracy can be measured by one or several metrics such as correlation ($r$) or mean squared error \cite{Morley2018}. The square of the correlation coefficient, median absolute error, mean absolute error, and mean absolute percent error were used to measure model performance (below we present a subset of these results). The MDA determines the overall importance of features in the final models; in the MDA analysis, the square of the correlation coefficient ($r^2$) is used to measure feature importance. 

In this work, we develop three random forest models: one using solar indices, one using FISM2 data, and a final model combining FISM2 and OMNI data \add[KRM]{as model features}. We determine the most important features for each model and compare the overall accuracy using the metrics described in the previous paragraph. However, before developing the three Random Forest models, a subset of data is used to determine the nominal hyperparameters or model settings. This optimization helps reduce the chance of overfitting in the models and ensures that the models generalize well, that is they work well on not only the training data but the test and out of sample (or validation) data as well. In Random Forests, the model hyperparameters include the number of trees in the forest, the number of bins (or leaves) a tree can end up with, the minimum number of samples in a bin before splitting and creating a new bin, and the number of features to consider when making a decision. To determine the optimal set of hyperparameters, a number of values is defined for each hyperparameter, each combination of hyperparameters is then looped through, and a Random Forest fits a subset of data and features. The combination of hyperparameters, which maximizes model performance, defines the optimal model setup. Here, the model hyperparameters are determined using a subset of the FISM2 and OMNI data as features, and GRACE-B density, model performance is measured using the mean absolute error. An excellent example of this framework is described and illustrated in \citeA{Bentley2018}.

Using the nominal set of hyperparameters, three Random Forest models are trained using the (i) solar indices, (ii) FISM2, and (iii) FISM2/OMNI as features. In each model, the target data \add[KRM]{(dependent variable)} is the GRACE-B neutral density, and select \add[KRM]{variables from each dataset are used as model features (independent variables)}. These features are determined using the correlation matrices shown in Figure \ref{fig:2} and the MDA feature importance. Each model's features and target data are separated into a train/test dataset using a 70/30 random split. Each model is trained on the training dataset. The MDA is then used to identify and remove features that add little to the model's overall performance. This helps ensure only key features are included and the final model is lean (e.g., reduce complexity). The test dataset and the out-of-sample GRACE-A and CHAMP datasets are used to validate model performance and ensure the model generalizes well to inputs it was not trained on, e.g., the model does not overfit to the training data. The training and out-of-sample datasets are also used to investigate model performance during quiet-times, storm-times, and select case studies. Finally, note that all datasets are taken the similar time period, 2002-2012.

\begin{figure}
    \includegraphics{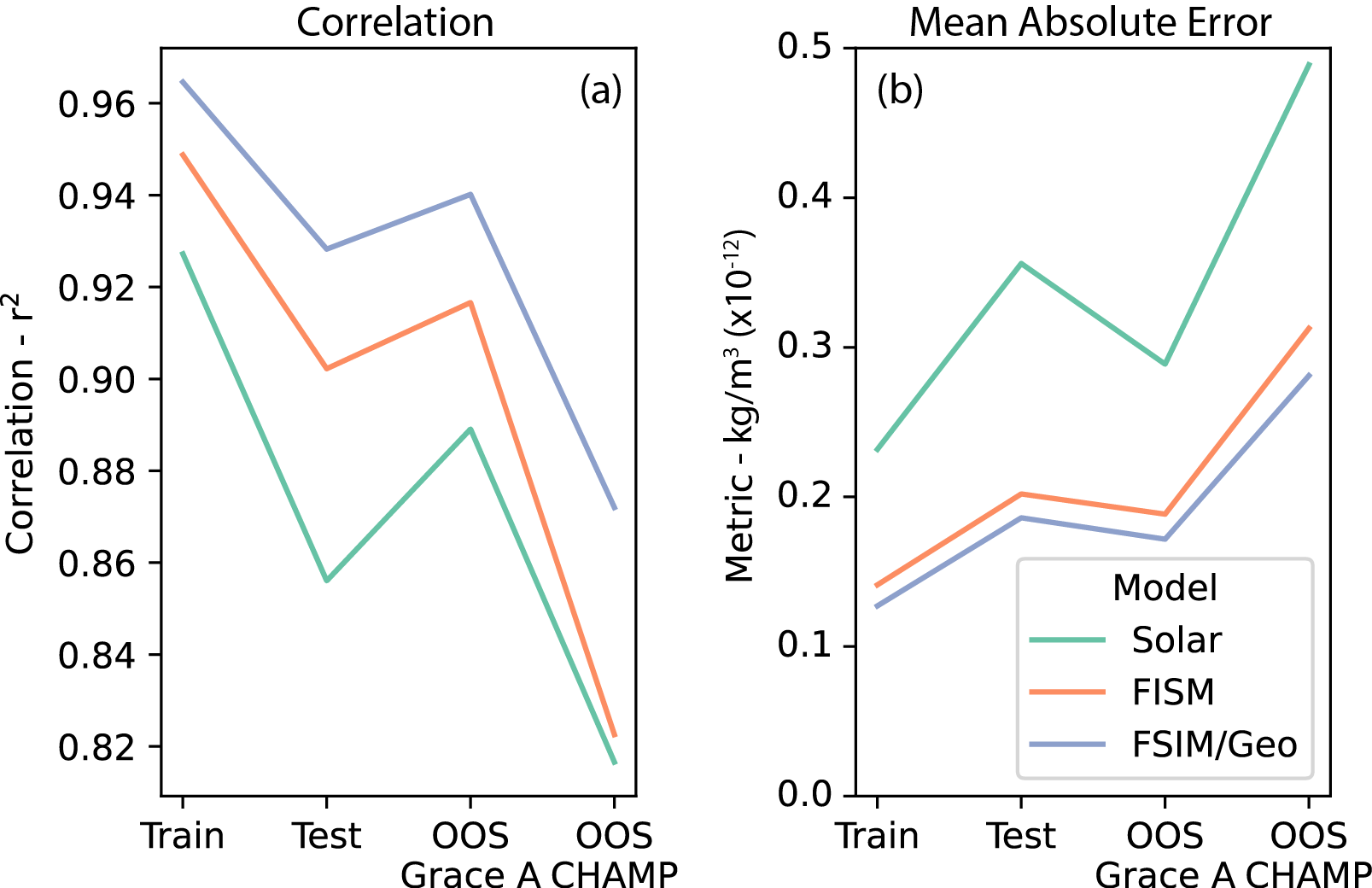}
    \caption{Metrics for the three random forest models as a function of dataset; train, test, and GRACE A and CHAMP out of sample (OOS). (a) The square of the correlation coefficients. (b) The mean absolute error.}
    \label{fig:3}
\end{figure}

Table \ref{tab:1} shows the final set of features for each of the three models. Magnetic local time (MLT) and latitude are the base features in each model. The solar indices model uses the hourly indices (but not the 81-day centered averages), the FISM2 model uses four key wavelengths, and the FISM2/OMNI model uses the same wavelengths as the FISM2 model along with the Sym-H and AE indices. Figure \ref{fig:3} shows the performance of each model as measured by the square of the correlations coefficient ($r^2$) and the mean absolute error. The metrics are calculated for the GRACE B train and test datasets and GRACE B and CHAMP the out-of-sample datasets. 

Overall, each model performs well with correlations between 0.82-0.96 and errors on the order of $0.1-0.5\times10^{-12} kg/m^3$. However, for each metric and dataset, the combined FISM2/GEO model performs best; it has the highest correlations and lowest errors. This is followed by the FISM2 model and then the solar indices model. In terms of datasets, unsurprisingly, the train dataset performs best, followed by the test and GRACE A datasets, which have very similar performances. The out-of-sample CHAMP dataset has the lowest correlation and largest errors of all datasets, though, for the FISM2/GEO model, the correlation remains high \~0.88 as compared to the FISM2 and solar models. The mean absolute error is also low $0.1\times10^{-12} kg/m^3$, about an order of magnitude smaller than typical values observed during geomagnetic storms ($>0.1\times10^{-11} kg/m^3$ Figure \ref{fig:1}). 

\begin{table}
 \caption{Features (independent variables) used in the random forest models. All features are treated the same, e.g., no weighting, in all models.}
 \begin{center}
 \begin{tabular}{c c c c}
 \hline
  Base Features (all models)  & Solar Features & FISM2 Features & FISM2/GEO Features  \\
 \hline
    MLT & F10, S10 & 1.30 $nm$ & FISM2 features \\
    $\cos(2\pi \times MLT/24)$  & Y10, M10 & 43.00 $nm$ & Sym-H \\
    $\sin(2\pi \times MLT/24)$  & F81, S81 & 85.55 $nm$ & AE \\
    latitude & Y81, M81 &  94.40 $nm$ & \\
 \hline
 \end{tabular}
 \end{center}
 \label{tab:1}
\end{table}

\subsection{Feature Importance}

As described in the previous section, the FISM2/OMNI model performs best, followed by the FIMS2 model and then the solar indices model. However, in this study, we also aim to quantify how adding features improves model performance, specifically the incorporation of geomagnetic data. In this section, we use the MDA to investigate the relative importance of model features in the FISM2 and FISM2/GEO models. The solar model is ignored as the common features are limited; more importantly, though, the previous section demonstrated that more accurate models could be developed using FISM2 and GEO indices.

\begin{figure}
    \includegraphics{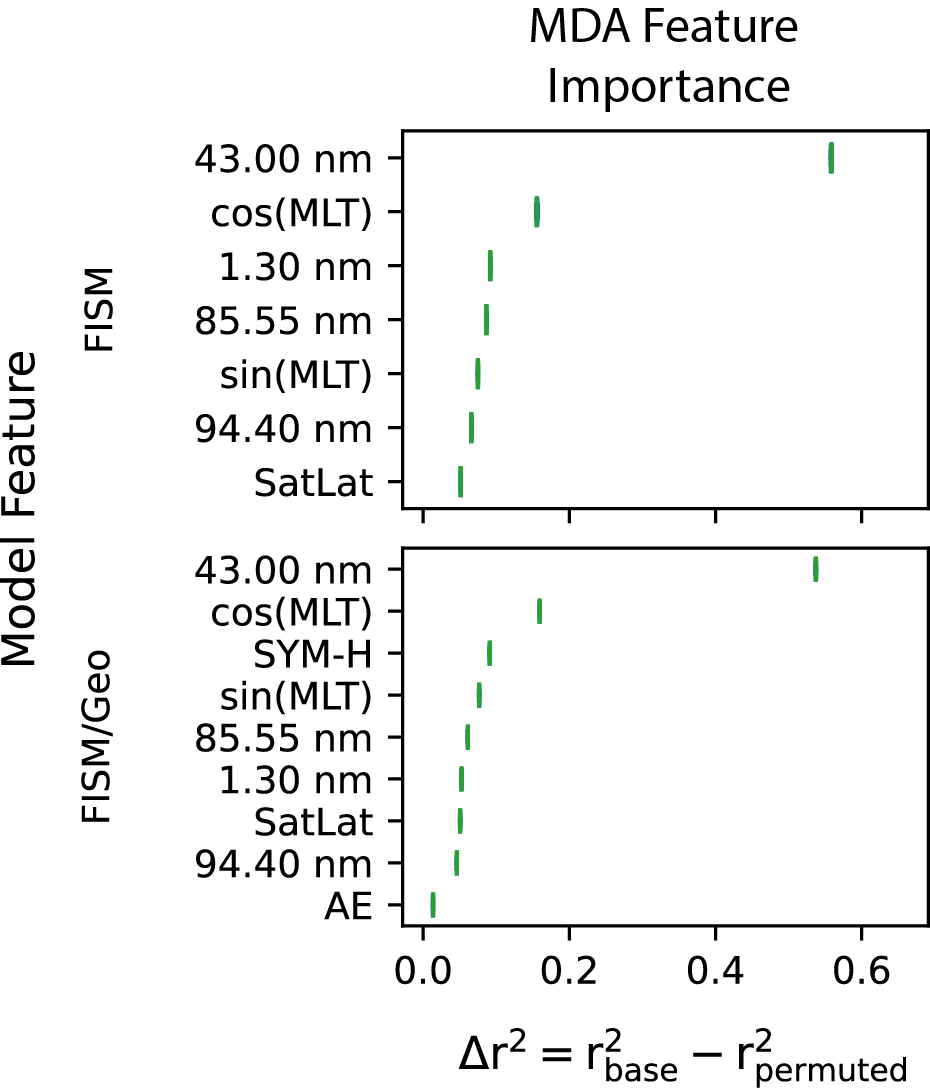}
    \caption{The MDA feature importance for the FISM (top) and FISM~/GEO models (bottom). In both panels, the features are ranked from most important (top) to least important (bottom).}
    \label{fig:4}
\end{figure}

Figure \ref{fig:4} shows the MDA values calculated for each feature in the FISM2 and FISM2/OMNI models using the out-of-sample CHAMP dataset. The out-of-sample dataset is used as it provides data that the model has not seen and allows for a more robust determination of feature importance. In Figure \ref{fig:4}, the MDA of each feature is calculated using a k-folds technique. A feature vector is randomly shuffled and used to predict the target value in a given model and calculate a measure (or score) of the models' accuracy. This is performed k-times for each feature. The MDA is then calculated as the difference between the base score and the average of the shuffled scores, MDA=$\delta r^2=r^2_{base} - \overline{r}^2_{shuffled}$. Here, we use the correlation squared as the model score, $r^2$.

Evident in Figure \ref{fig:4} is that in both the FISM2 and FISM2/OMNI models, the most important feature is the 43.00 nm spectral band. Shuffling of this feature reduces $r^2$ by nearly 0.6. This is followed by magnetic local time as characterized by $\cos(MLT)$, shuffling of which reduces $r^2$ by about 0.2 in both models. In the FISM model, this is followed by the remaining spectral bands and $\sin(MLT)$. In the FISM2/GEO model, the next important feature is Sym-H, followed by the $\sin(MLT)$, the remaining spectral bands, and AE. Note in both models, the satellite latitude (SatLat) contributes very little to the overall importance of each model, and in the FISM2/GEO model, AE is the least important feature. This suggests that while atmospheric density is highly dependent on magnetic local time (e.g., day vs night) and Sym-H, there is less variation with latitude and possibly nightside activity (as measured by AE), indicating that global phenomena such as geomagnetic storms may play a more important role in enhancing atmospheric density than more localized phenomena such as substorms. In the next section, we investigate model performance during select case studies and statistically over both quiet- and storm times.  

\subsection{Case and Statistical Studies}

\begin{figure}
    \includegraphics{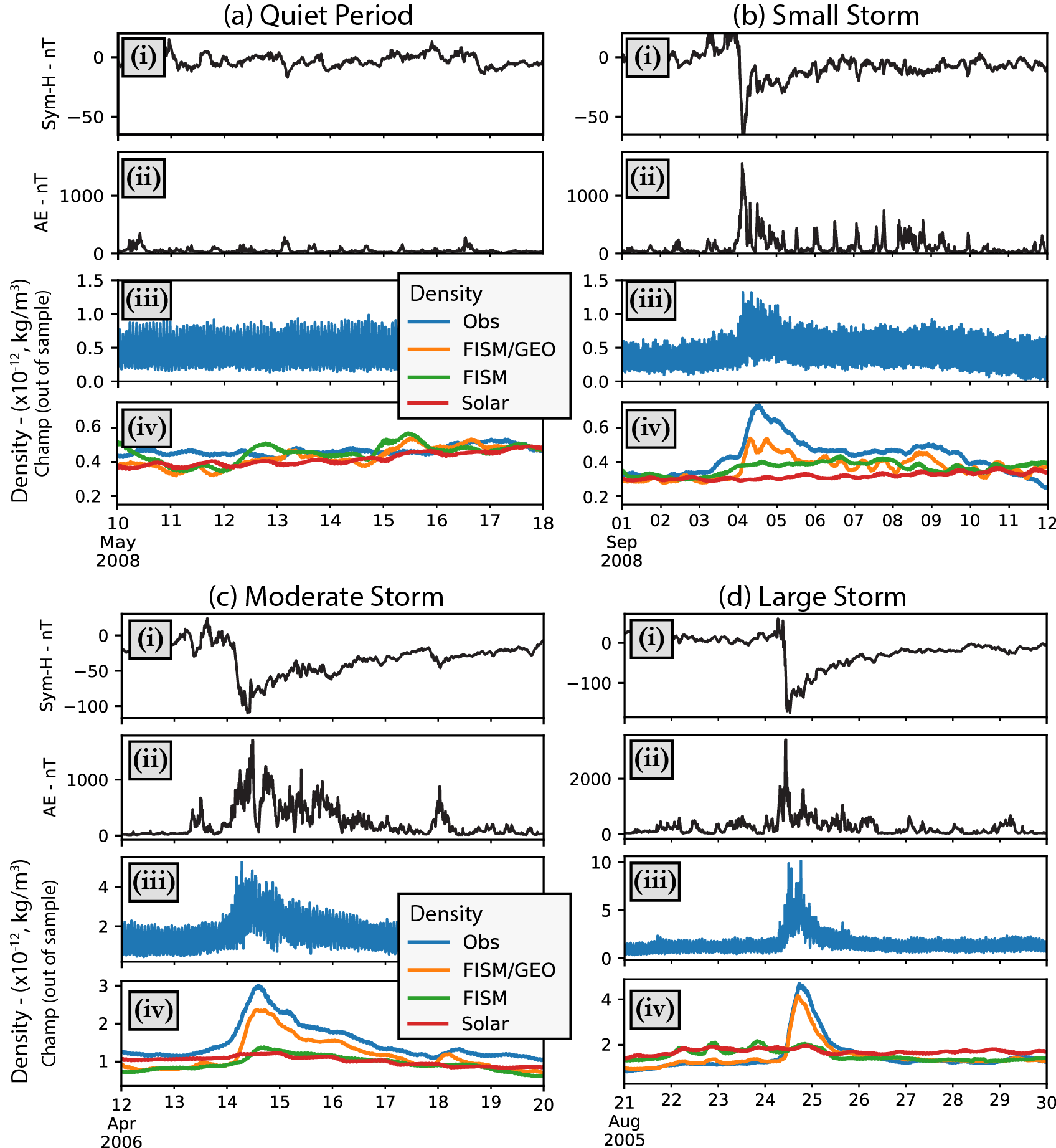}
    \caption{Select case studies comparing observed density from the out-of-sample CHAMP dataset and that derived from the three random forest models during (a) geomagnetically quiet period, (c) small geomagnetic storm, (c) moderate geomagnetic storm, and (d) large geomagnetic storm. From top to bottom, each panel shows (i) Sym-H, (ii) AE, (iii) Champ observed density, and (iv) model-data comparison. In panel (iv) the densities have been averaged with a 90 minute rolling window to make comparisons easier and highlight the background change in density during each storm. Observed densities are shown in blue, FISM/GEO modeled densities in orange, FISM modeled densities in green, and solar-modeled densities in red.} 
    \label{fig:5}
\end{figure}

This section reviews model results during a select quiet period and three increasingly severe geomagnetic storms (as measured by Sym-H). This is followed by a statistical investigation of model residuals as a function of quiet- and storm-times. Figure \ref{fig:5} shows the four case studies (a) a geomagnetically quiet period, (b) a small geomagnetic storm, (c) a moderate storm, and (d) a large storm. Each event shows the same four panels, from top to bottom: (i) Sym-H, (ii) AE, (iii) Champ density normalized to 400 km, and (iv) comparison of Champ and the Random forest modeled densities along the Champ trajectory. Overall, the observed density is highly variable (panels iii), though it typically follows a background trend, especially during storms (b-d). The quiet-time densities are typically low, $ < 10^{-12} kg/m^3 $, and increase with the geomagnetic storm intensity by a factor of 2 during the small storm and \~10 during the large storm. The storm-time density enhancements peak during the main phase of each storm as Sym-H decreases and AE activity increases. As the storm recovers and Sym-H and AE trend toward zero, the density enhancements decay toward quiet-time levels.  

Panel (iv) shows the out-of-sample data-model comparisons. Note that the high variability in the density time series makes any visual comparison difficult; thus, the modeled and observed densities are smoothed with a 90-minute rolling average - the approximate orbital period for a low Earth orbit satellite. Further, we compare the out-of-sample model results and data to get a better idea of the true performance of the three models as models perform better on datasets used in training. Evident in panel (iv) is that during geomagnetic storms (b-c), the combined solar-geomagnetic model (FISM/GEO - orange) performs best, capturing the rapid density enhancement and slow decay density as well as the amplitude of the density enhancement (especially during the moderate and large storm). The two solar models (FISM - green, Solar - red) perform poorly during storms as they are unable to capture the background trend in the variation in atmospheric density during storms. During quiet times (a), all three models generally capture the background variation.

\begin{figure}
    \includegraphics{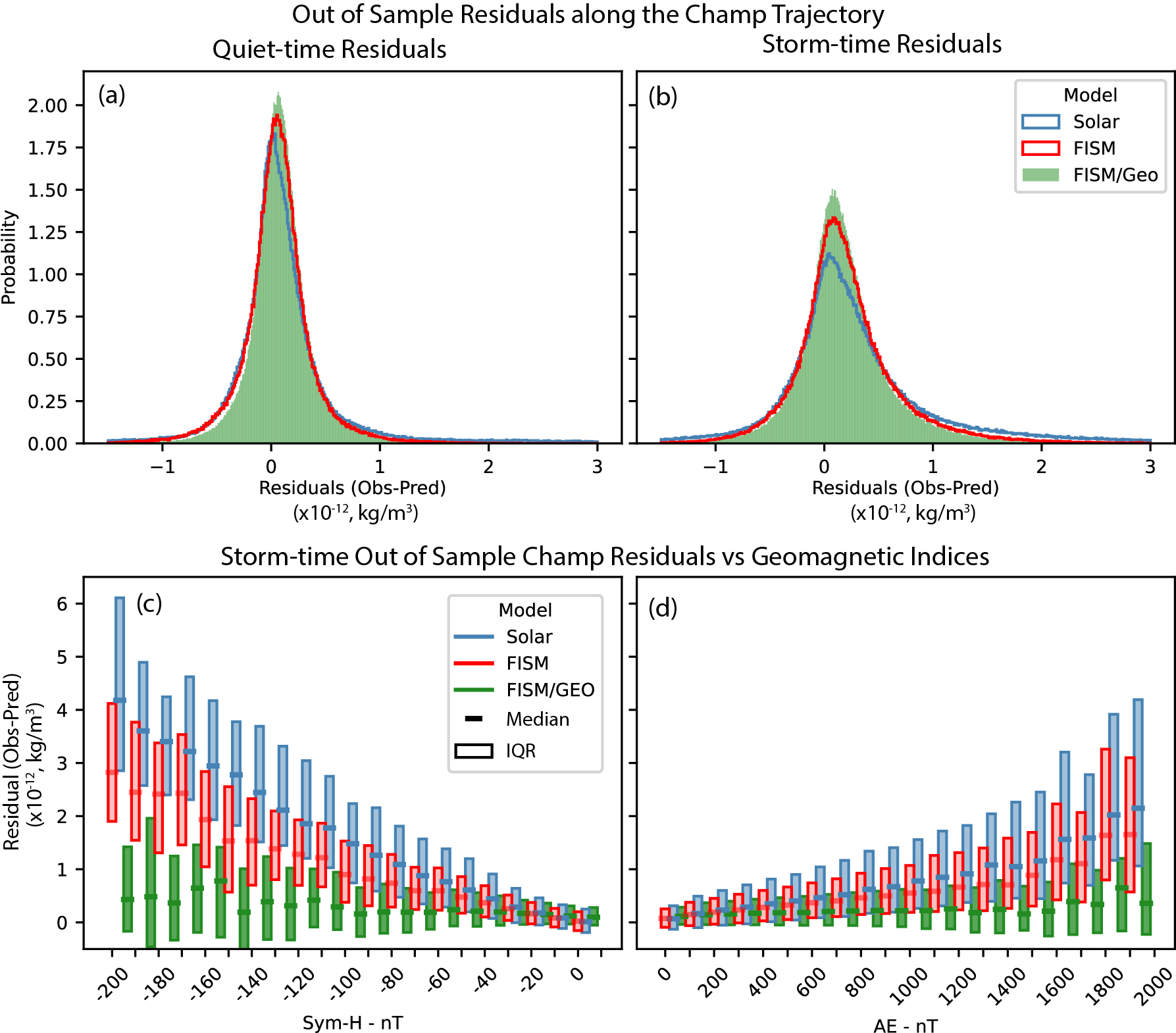}
    \caption{Out of sample Champ residuals as a function of quiet-time and storm-times. Panels (a) and (b) show the probability distributions of the residuals during quiet times and storm times, respectively. Panels (c) and (d) further break down the storm-time residuals as a function of geomagnetic activity measured by Sym-H and AE. In (c) and (d), the residuals are binned by geomagnetic activity, and the median of the residuals is plotted as a solid line and the interquartile range (IQR) as a box for each bin. The top of the boxes represents the third quartile, and the bottom of the boxes represents the first quartile. In panel (c), Sym-H is binned in 10 nT increments. In panel (d), AE is binned in 100 nT increments. In each panel, the Solar model is shown in blue, FISM in red, and FISM/GEO in green.}
    \label{fig:6}
\end{figure}

Figure \ref{fig:6} shows a statistical analysis of the out-of-sample CHAMP residuals for each model as a function of quiet- and storm-times. Panels (a) and (b) show probability distributions of the residuals during geomagnetic quiet periods and storms, respectively (note panels a and b share a y-axis). During quiet-times (a) all three models perform well, the residuals are peaked around zero with similar probability distributions and a slight bias toward the models underestimating the actual density (positive residuals). However, during storm-times (b) the FISM/GEO model (green) performs better than the FISM (red) and Solar (blue) models. The FISM/GEO model highly peaked around zero, while the FISM and Solar models have smaller peaks around zero and larger tails toward larger positive and negative residuals. Again, during storms, each model underestimates the observed density (shift toward positive residuals).

Figure \ref{fig:6} (c) and (d) show the storm-time model residuals as a function of geomagnetic activity as measured by Sym-H and AE, respectively (note panels c and d share a y-axis). Here, the residuals are binned by geomagnetic activity, and the residuals are plotted similarly to a box-and-whisker plot: the median is shown by the solid line, and the boxes show the inter-quartile range (bottom of the box is the first quartile, and the top of the box is the third quartile). These panels very clearly highlight the importance of storms and geomagnetic activity in quantifying the dynamics of atmospheric density. In particular, as geomagnetic activity increases during storms and Sym-H becomes more negative and AE more positive, the residuals in the Solar (blue) and FISM (green) models rapidly increase with very large interquartile ranges. However, the residuals of the combined FISM/GEO model remain low, with the median $<$1 and upper quartile $<$2. Overall, as storm activity increases, the solar models can have errors up to \~8 times larger than the model that combines both solar and geomagnetic data and, further, has significant spread as a function of geomagnetic activity such that model accuracy decreases with increasing geomagnetic activity. These differences agree with the case studies highlighted in Figure \ref{fig:5}.   

\section{Discusion and Conclusions}

As the number of satellites in low Earth orbit increases, it becomes increasingly important to track and predict their orbits accurately. This is especially true with the advent of mega-constellations like Starlink, Globalstar, OneWeb, Telesat Lightspeed, and Project Kuiper, which are each composed of hundreds to thousands of small satellites. Additionally, countries around the world have plans for LEO mega constellations; it is expected in the next decade, that nearly 100,000 constellation satellites will be launched \cite{Zhang2022}, a number which exceeds the total of satellites launched in the first half century of the space age by over a factor of 16. While these satellites have important societal benefits, including increasing access to high-speed internet in remote areas, the increase in LEO satellites has exponentially increased the number of close encounters between satellites, which in-turn increases the chances of collisions. Mitigating collisions and monitoring/forecasting close encounters requires accurate knowledge and prediction of satellite trajectories. 

Of all the factors controlling satellite trajectories, drag is the key factor in achieving high-fidelity orbit determination. The biggest factor in determining satellite drag is atmospheric density and its spatial and temporal variations. The dynamics of atmospheric density are driven by a combination of external forcing from the Sun \cite{Lilensten2008} and the magnetosphere \cite{Knipp2004} and internal processes in the lower atmosphere \cite{Liu2016}. During quiet geomagnetic conditions, the dominant source of variations in atmospheric density is solar activity; however, during geomagnetic storms, energy input from the magnetosphere can lead to rapid changes in atmospheric density ranging from 50-800\% \cite{Forbes1996, Liu2016, Oliveira2019}. Here we investigated the dynamics of atmospheric density observed by GRACE-B during geomagnetic quiet periods (quiet-times) and geomagnetic storms (storm-time) as function of solar and geomagnetic activity and developed, tested, and contrasted three Random Forest models of atmospheric density based on low-cadence solar spectral irradiance (those used in JB2008), high-cadence solar irradiance (from the FISM2 dataset), and combined high-cadence solar irradiance and geomagnetic activity (from the Omni dataset). The models were validated for train and test datasets and two out-of-sample datasets, GRACE-A and CHAMP. The CHAMP out-of-sample dataset was further used to test model performance during select case studies and statistically investigate model residuals during storms as a function of geomagnetic activity.

As described above, atmospheric density can vary considerably during geomagnetic storms \cite{Forbes1996, Liu2016, Oliveira2019}. While it is known that this is the result of energy input from the magnetosphere, little work exists quantifying storm-time atmospheric density and investigating the difference between quiet- and storm-times. Recent case studies examining several geomagnetic storms with and without accompanying flares proposed that the flares only contributed a minor amount to the resulting atmospheric disturbances \cite{Qian2020}. The work presented here examines the relative contributions of geomagnetic inputs and solar irradiance over a large statistical dataset for both storm- and quiet-time conditions. Figure \ref{fig:1}, shows the PDF and CDF of atmospheric density as a function of storm-times, storm phase, and quiet-times. Figure \ref{fig:2} shows the correlations between the low-cadence solar irradiance indices (a), the higher-cadence FISM2 irradiance dataset (b), and solar wind and geomagnetic activity (c) with atmospheric density. Taken together, Figures \ref{fig:1} and \ref{fig:2} demonstrate that while solar driving is a key factor in the baseline or long term dynamics of atmospheric density, fast changes and magnetospheric processes are a vital component storm-time atmospheric dynamics as evidenced by the correlations of atmospheric density during storms with the high cadence FISM2 and OMNI datasets.  

To explore this in more detail and develop a clearer understanding of the dynamics and drivers of storm-time atmospheric density, and more importantly, develop a robust model of atmospheric density capable of capturing storm-time dynamics we subsequently developed Random Forest machine learning models of neutral density. Three models were developed which shared base features (column 1 of Table \ref{tab:1}) but were trained on the low-cadence solar wind irradiance indices, that high-cadence FISM2 solar irradiance bands, and a final model which combines features from the FISM2 solar irradiance bands and OMNI geomagnetic data. The features from each model were selected based on the correlations shown in Figure \ref{fig:2} and feature importance as measured by the mean decrease in accuracy (MDA). The selected features for each model are summarized in columns 2-4 of Table \ref{tab:1}, respectively. The models were compared using accepted and appropriate metrics on the train and test datasets (Grace-B observations) and two out-of-sample datasets, GRACE-A and CHAMP. In all cases the model combining high-cadence data from the FISM2 and OMNI datasets performed best, followed by the FISM2 model, and then by slow cadence solar indices model (c.f., Figure \ref{fig:3}). Figure \ref{fig:4}, further investigated the relative importance of model features in the two high-cadence models (FISM and FISM/Geo) using the mean decrease in accuracy (MDA). This analysis helps to quantify the relative importance of solar and geomagnetic drivers of atmospheric density dynamics. In both high-cadence models, the 43.00 nm solar irradiance waveband is the most important feature, followed by position; however, in the FISM/Geo model the third most important feature is Sym-H, often a measure of the strength of geomagnetic storms. This supports the conclusion that magnetospheric processes are a fundamental driver of storm-time atmospheric dynamics. 

The metrics and MDA analysis in Figures \ref{fig:3} and \ref{fig:4} provide important insight into density dynamics; however, it is important to note that neither of these separated the data into quiet- or storm-times. In order to understand and quantify the importance of magnetospheric forcing in atmospheric dynamics, Figures \ref{fig:5} and \ref{fig:6} present an analysis of select case studies and a detailed investigation of model residuals for each of the three models as a function of quiet-times, storm-times, and geomagnetic activity and using only the out-of-sample CHAMP data. In the select case studies presented in Figure \ref{fig:5} all three models do well during quiet times (panel a); however, as geomagnetic activity increases, both solar models do poorly while the FISM/Geo model does very well, capturing the shape and time scale of the enhancement though slightly underestimating the peak. Figure \ref{fig:6} further quantifies the residuals (observations-predictions) as a function of quiet-times (a), storm-times (b), and geomagnetic activity (c and d). Similar to the case studies all three models do well during quiet-times while the FISM/Geo model does best during storm-times. Most interestingly, panels (c) and (d) demonstrate that as geomagnetic activity increases the errors (or residuals) in the two solar models increase at an almost exponential rate. In contrast, the errors in the FISM/Geo model remain flat. Overall, Figures \ref{fig:5} and \ref{fig:6}, very clearly demonstrate the importance of magnetospheric forcing during storms in atmospheric density, and highlight the necessity of including and considering magnetospheric dynamics in atmospheric models. 

As noted prior, several studies have demonstrated the variation of atmospheric density during storms. Thus, it is not surprising that the solar models are not able to capture the dynamics of atmospheric density during storms. What is important to note is that is that during storms the initial enhancement of density is rapid (<24 hours), and the decay can occur over several days. In order to capture such dynamics it is imperative to user higher cadence datasets such as the FISM2 and OMNI datasets, slower observations, especially those on the order of days will be unable to capture such variations in the atmospheric density. These enhancements and decay timescales have important implications when quantifying the performance of models, those here, as well as models in general. If one considers only storm-times or storm phase one can incorrectly assert that in general model errors are small and well behaved, for example, when calculating the mean or median error (e.g., Figure \ref{fig:6} b). However, if you consider the full time history during storms (Figure \ref{fig:5}) or errors as a function of geomagnetic activity (Figure \ref{fig:6} c and d), it becomes very clear that solar indices alone are not able to capture the storm-time dynamics of atmospheric density. This is quite surprising and very clearly illustrated in Figure \ref{fig:6}. Errors computed on the out-of-sample data increase nearly exponentially with geomagnetic activity in the solar models. In contrast, the errors in the combined solar-geomagnetic model remain flat as a function of geomagnetic activity. 

Overall, the work in this paper has demonstrated and quantified the importance of magnetospheric process and geomagnetic activity in storm-time atmospheric density dynamics. Future work will investigate the two-dimensional distribution and variation in atmospheric density during storms as simulated by the FISM/Geo model, introducing additional datasets and features in the FISM/Geo model to improve performance, expanding the FISM/Geo model to include altitude, and finally working toward transitioning this novel research model to operations and developing outputs of use to key stakeholders.


\section{Open Research}

All data used in this manuscrip is freely avialable online. The solar indices were obtained from Space Environment Technologies via https://sol.spacenvironment.net/JB2008/. The FISM2 dataset was obtained from the LASP Interactive Solar IRrdiance Datacenter (LISIRD) \cite{LISIRD} via https://lasp.colorado.edu/lisird/data. The solar wind and geomagnetic data were obtained from NASA Space Physics Data Facility (SPDF) via https://omniweb.gsfc.nasa.gov/. The neutral densities were obtained from the University of Colorado Boulder Space Weather Data Portal \cite{SWxTREC} via https://lasp.colorado.edu/space-weather-portal/. \add[KRM]{Scikit-learn was used to develop the Randfom Forest models.}


\acknowledgments

The Space Precipitation Impacts (SPI) Heliophysics internal funding model partially funded K.R.M., A.J.H. K. G-S., J. K., and J. S. V. L. Was funded through the Helioanaytics summer internship.


%
%



\bibliography{SatDragConcept}

%
%
%
%
%

\end{document}